\documentclass[12pt]{article}
\usepackage[english]{babel}
\usepackage{amssymb,cite}
\usepackage{epsfig}

\def\starup#1{\mbox{$\raise1.8ex\hbox{$*$} \kern-.7em#1$}}
\def\krup#1{\mbox{$\raise1.8ex\hbox{$+$} \kern-.7em#1$}}

\begin{document}
\title{Bounds \\ 
       on scalar leptoquark and scalar gluon masses \\
       from current data on S, T, U } 
\author{A.D.~Smirnov\thanks{E-mail: asmirnov@univ.uniyar.ac.ru}\\
{\small\it Division of Theoretical Physics, Department of Physics,}\\
{\small\it Yaroslavl State University, Sovietskaya 14,}\\
{\small\it 150000 Yaroslavl, Russia.}}
\date{}
\maketitle

\begin{abstract}
The contributions into radiative correction  
 pa\-ra\-me\-ters S, T, U   
from scalar leptoquark and scalar gluon doublets 
are investigated in the minimal four color symmetry model. 
It is shown that 
the existence of the relatively light 
scalar leptoquarks and scalar gluons  
(with masses of order of 1 TeV or less) 
is consistent with the current experimental 
data on S, T, U and the more light particles 
(with masses below 400 GeV) improve the fit.  
In particular, the lightest scalar leptoquarks 
with masses below 300 GeV are consistent  
with the current data on S, T, U at 
$\chi^2 < 3.1(3.2)$ 
for  
 $m_H=115(300)\,\,GeV$ 
in comparison with 
$\chi^2 = 3.5(5.0)$ 
in the Standard Model.  
The lightest scalar gluon in this case is expected 
to lie below $ 850(720) \,\,\,GeV.$
The possible significance of such particles 
in the t-quark physics at LHC is emphasized 
and their effect on the gauge coupling constant 
unification is briefly discussed.  
\end{abstract}


\section{Introduction}

By this talk I would like to call an attention to 
the possibility of scalar leptoquarks and scalar gluons 
relating to the four color quark-lepton symmetry 
 to be relatively light.  

Apparently each of us is interested in a question 
what kind of the new physics beyond the Standard Model (SM) 
can exist at more high energies and how this new physics 
can manifest itself at the energies of the present and 
future colliders. 
One of the possible variants
of such new physics can  be the variant induced 
by the possible four color symmetry \cite{PS} 
between quarks and leptons.
The immediate consequence of this symmetry is the prediction 
of the new gauge particles -- 
vector leptoquarks with the masses of order of the mass 
scale $M_c$ of the four-color symmetry breaking. 
In dependence on the model 
the lower limit on $M_c$ can vary from 
 $M_c\sim 10^{12}\,\,GeV$ \cite{EF} 
or  $M_c\sim 10^5 - 10^6 \,\,GeV$  \cite{SS} 
in GUT models with the four-color symmetry as an intermediate stage 
of symmetry breaking to   
 $M_c\sim 1000\,\,TeV$ \cite{V} 
or to  $M_c\sim 100\,\,TeV$  
or less in the models 
regarding the four-color symmetry as a primery symmetry 
\cite{V,AD1,AD2,RF1,RF2,YF,BL} 

It should be noted however that 
the four-color symmetry can manifest itself not only by the vector 
leptoquarks but also due to the new scalar particles such as 
scalar leptoquarks \cite{BVR,HR} or scalar gluons \cite{AD1,AD2}. 
In particular, 
in addition to the vector leptoquarks 
the four color symmetry with the Higgs mechanism 
of splitting the masses of quarks and leptons 
(MQLS-model, \cite{AD1,AD2}) predicts \cite{PovSm1} 
the scalar leptoquarks and the scalar gluons 
of the doublet structure under the electroweak 
$ SU_L(2) $-group.  
In this approach these doublets are responsible 
for splitting the masses of quarks from those of 
leptons and they are the partners of the standard 
Higgs doublet. 
What can we say about the masses of these scalar 
doublets? 

These particles could manifest themselves through 
the radiative corrections. As is known if the new particles 
are relatively heavy ($m_{new} \gg m_Z$) their contributions 
into electroweak corrections can be approximately accounted 
by the formalism of the  $ S, T, U- $ parameters of 
Peskin and Takeuchi \cite{PT}. 
The current experimental data  
\cite{PDG01} 
give the next values of 
$ S,\,T,\,U $  
which can be induced by a new physics  
\begin{eqnarray}
S_{new}^{exp}&=& -0.03\pm 0.11\,\,(-0.08),
\nonumber\\
T_{new}^{exp}&=& -0.02\pm 0.13\,\,(+0.09),
\label{eq:stue} \\
U_{new}^{exp}&=& \;\;\; 0.24\pm 0.13\,\,(+0.01), 
\nonumber
\end{eqnarray}
where the central values assume $m_H = 115\,\,GeV$ and the change for 
$m_H = 300\,\,GeV$ is shown in parentheses. 
The $ S,\,T,\,U - $ parameters are normalized so that 
in SM they are equal to zero ($ S_{SM}=T_{SM}=U_{SM}=0 $). 

In this talk I would like to discuss the bounds on the masses 
of the scalar leptoquark and scalar gluon doublets 
wich are imposed by the current 
$ S,\,T,\,U $ 
data 
(\ref{eq:stue}). 

\section{Brief description of the MQLS model}

The MQLS-model to be used here is based on the
$SU_V(4) \times SU_L(2)\times U_R(1)$-group 
and predicts the new 
gauge particles ( vector leptoquarks 
$V{\alpha \mu}^\pm$ 
with electric charge ${\pm}$2/3 and an extra neutral 
$Z'-$ 
boson ) as well as the new scalar ones \cite{AD1,AD2}.

The scalar sector of the model contains in general the four multiplets 
$
   \Phi_A^{(1)},\,\,
   \Phi_a^{(2)},\,\, 
  \Phi_{i,a}^{(3)}, \,\,
  \Phi_i^{(4)}, \,\,
$
transforming according to the  
(4,1,1), (1,2,1), (15,2,1), (15,1,0) representations of the
$SU_V(4) \times SU_L(2)\times U_R(1)$-group  
and having the vacuum expectation values 
$\eta_1$, $\eta_2$, $\eta_3$, $\eta_4$ 
respectively. 
Here 
$ A=1,2,3,4$ and $i=1,2\,...\,15$ are the $SU_V(4)$ indexes and 
$ a=1,2$ is the $SU_L(2)$ one. 

The scalar leptoquark and scalar gluon doublets  
$ S_{a\alpha}^{(\pm)}$ and $F_{ja}$ 
to be discussed here belong to the (15,2,1)-multiplet 
$\Phi_{i,a}^{(3)}$.  
This multiplet together with (1,2,1)-multiplet 
$\Phi_{a}^{(2)}$ 
generates the fermion masses by Higgs mechanism 
and splits the masses of quarks from those of leptons.  

Below we consider the scalar leptoquark doublets 
in the case of the 
the simplest scalar leptoquark mixing 
and with neglect of the small parameter 
$\xi^2= \frac23g_4^2\eta_3^2/m_V^2 \ll 1 $ 
of the model. 
In this case the scalar leptoquark doublets can be written as 
\begin{eqnarray}
S^{(+)} = 
\left ( \begin{array}{c}
S_1^{(+)}\\
c\,\, S_1 + s\,\, S_2
\end{array} \right )  ,
S^{(-)} = 
\left ( \begin{array}{c}
S_1^{(-)}\\
-s\,\, \starup{S_1} + c\,\, \starup{S_2}
\end{array} \right )  ,
\label{eq:SpSm}
\end{eqnarray}
where $S_1,\,\,S_2$ are the mass eigen states of 
the scalar leptoquarks with electric charge 2/3 and 
$c=\cos\theta,\,\,s=\sin\theta,\,\,\theta $ 
is the scalar leptoquark mixing angle. 

The scalar gluon doublets in general case 
can be written as  
\begin{eqnarray}
F_j = 
\left ( \begin{array}{c}
F_{1j}\\
(\phi_{1j}+i\phi_{2j})/\sqrt{2}
\end{array} \right )  ,
\label{eq:Fj}
\end{eqnarray}
where the charged fields $F_{1j}$ and the neutral fields 
$\phi_{1j},\phi_{2j},j=1,2,...,8$ are the mass eigen state fields 
(in general case the real and imaginary parts $\phi_{1j},\phi_{2j}$ 
of the down component of the doublet $F_j$ can be splitted in the mass).  

The contributions 
$S^{(LQ)}$, $T^{(LQ)}$, $U^{(LQ)}$ 
and
$S^{(F)}$, $T^{(F)}$, $U^{(F)}$ 
into 
$S,\,T,\,U$ 
from the scalar leptoquark and scalar gluon doublets 
have been calculated and analysed in  
ref.\cite{AD3,AD4,AD5}. In the case of the scalar doublets 
of the forms   
(\ref{eq:SpSm}), 
(\ref{eq:Fj})
the resulting contributions 
$S=S^{(LQ)}+S^{(F)}$,  
$T=T^{(LQ)}+T^{(F)}$,   
$U=U^{(LQ)}+U^{(F)}$ 
depend on seven masses 
and on the mixing angle $\theta$.  
The contributions  
 $T^{(LQ)}$ and $U^{(LQ)}$ 
from the scalar leptoquark doublets are not positive definite 
due to the $S_1-S_2$- mixing 
and can be negative if 
$ m_{S_1^{(+)}}, m_{S_1^{(-)}} $  
are between $m_{S_1}$ and $m_{S_1}$  
and the contributions $T^{(F)}$ and $U^{(F)}$ are also not positive definite 
and they are negative if $m_{F_1}$ is between $m_{\phi_1}$ and 
$m_{\phi_2}$. 

The masses of the scalar leptoquark and scalar gluon doublets 
are generated 
by the Higgs mechanism of the symmetry breaking 
from the scalar potential 
$ V(\Phi^{(SM)}, S^{(+)}),S^{(-)},F) $ 
including the interactions of these doublets with 
the standard Higgs doublet. 
In this case we obtain some relations between 
the masses of scalar particles and the coupling 
constants of the scalar potential. As a result 
we have the new eight independent parameters of the model  
\begin{eqnarray}
 m_{S_1^{(+)}}, m_{S_1^{(-)}}, m_{F_1}, \gamma_{+}, \gamma_{-}, \delta_S, 
\gamma_F, \delta_F,  
\label{eq:par}
\end{eqnarray} 
where 
$ \gamma_{+}, \gamma_{-}, \delta_S, 
\gamma_F, \delta_F $ 
are the coupling constants describing the interactions 
of the scalar leptoquark and scalar gluon doublets 
with the standard Higgs doublet.   

For the stability of the vacuum the coupling constants 
in the scalar potential are supposed below to satisfy 
some conditions ensuring the positiveness of the  
scalar potential   
\begin{eqnarray}
 V(\Phi^{(SM)}, S^{(+)}),S^{(-)},F) > 0.  
\label{eq:Vg0}
\end{eqnarray}

For validity 
of the perturbation theory the coupling constants 
in the scalar potential 
cannot be too large.  
We suppose below that all the coupling constants  
in the scalar potential 
do not exceed some maximal value 
$\lambda_{max}$ 
ensuring the validity of perturbation theory.
In the further numerical analysis we restrict ourselves
by the values of  
$\lambda_{max} $
from the region
$\lambda_{max} = 1.0 - 4.0 $ 
which give the reasonable values of the perturbation 
theory expansion parameter of order 
$\lambda_{max}/4\pi = 0.1 - 0.3 $. 

\section{Fit of the model parameters and discussion}

Varying the fitting parameters 
(\ref{eq:par}) 
we minimize
$\chi^2$ 
defined as 
$$
\chi^2=\frac{(S-S_{new}^{exp})^2}{(\Delta S)^2}+
\frac{(T-T_{new}^{exp})^2}{(\Delta T)^2}+
\frac{(U-U_{new}^{exp})^2}{(\Delta U)^2},
$$
where 
 $\Delta S,\Delta T, \Delta U$ 
are  the experimental errors in 
(\ref{eq:stue}). 

To clear up the possible effect of the scalar leptoquark 
and scalar gluon doublets on   
$S,\,T,\,U$ 
we vary the masses of these particles so that  
\begin{eqnarray}
m_{S_1}, m_{S_2}, 
m_{S_1^{(\pm)}}, 
m_{F_1}, 
m_{\phi_1}, 
m_{\phi_2} \ge m^{lower}_{scalar} ,
\label{eq:mlower}
\end{eqnarray} 
where 
$m^{lower}_{scalar}$ 
is a lower limit on the masses of these particles. 
 After minimization of 
$\chi^2$ 
under condition (\ref{eq:mlower})
we have analysed the dependence of 
$\chi^2_{min}$ 
on this lower limit 
$m^{lower}_{scalar}$ 
and on upper limit  
$\lambda_{max}$ 
on the coupling constants of the scalar potential. 

\begin{figure}[htb]
\centerline{\psfig{file=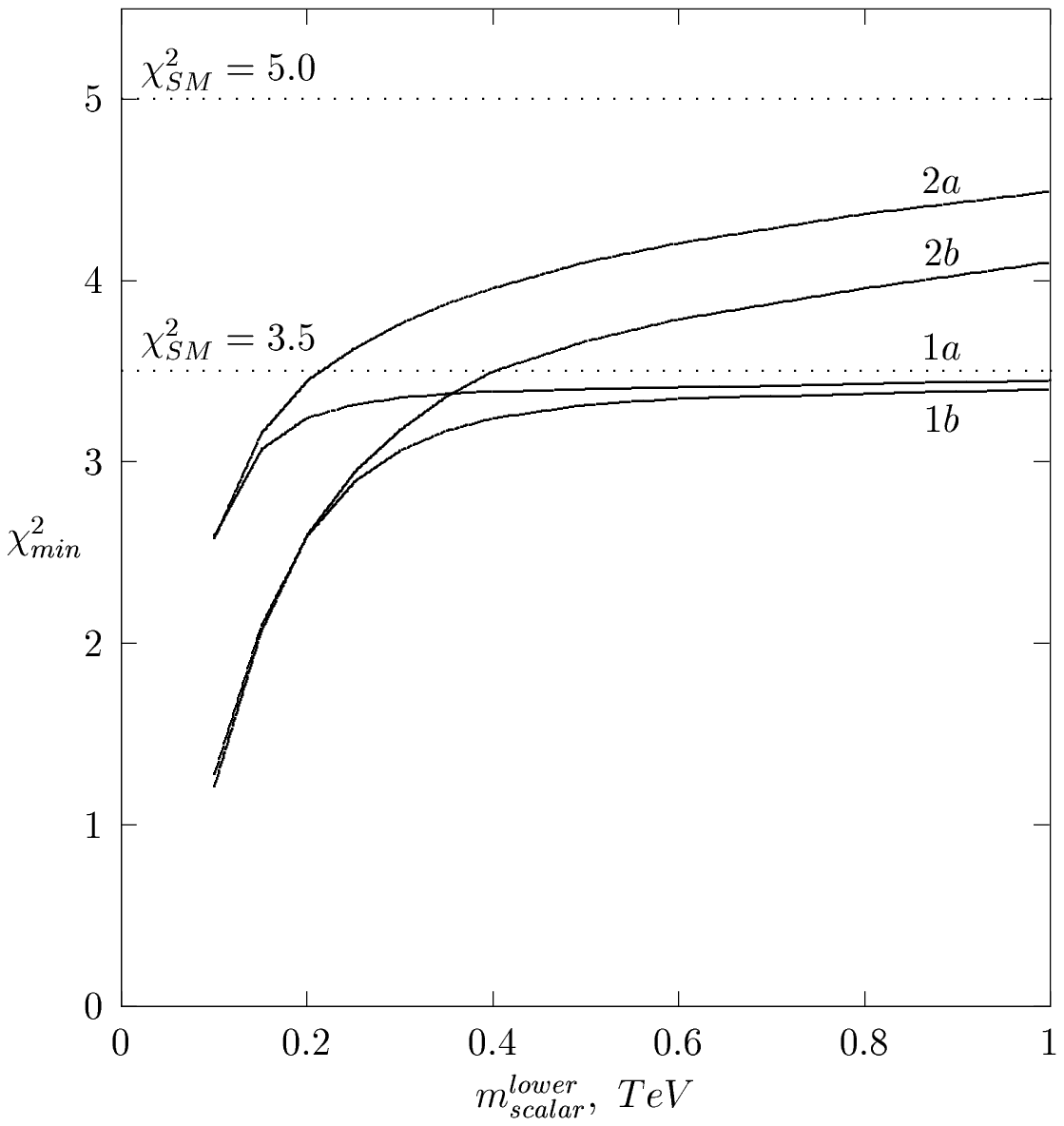,height=90mm}}
\caption{ $\chi^2_{min}( m^{lower}_{scalar},\lambda_{max})$ 
        as a function of the lower limit 
        $ m^{lower}_{scalar} $  
        on the masses of the scalar particles  
        for $m_H=115\,\,GeV$ (1) 
        and for $m_H=300\,\,GeV$~(2) 
        at $\lambda_{max}=1.0 (a) $ and  
        $\lambda_{max}=4.0 (b) $.  } 
\end{figure}

The Fig.1 shows 
$\chi^2_{min}( m^{lower}_{scalar},\lambda_{max})$ 
as a function of the lower limit 
$ m^{lower}_{scalar} $  
for $m_H=115\,\,GeV$ (the curves 1) 
and for $m_H=300\,\,GeV$ (the curves 2) 
at $\lambda_{max}=1.0(4.0)$ 
(the curves $a(b)$)
for the case without scalar leptoquark mixing 
($\theta=0$, this case is slightly preferred by 
$\chi^2$ minimum ). 
The horizontal lines denote  
$\chi^2_{SM} = 3.5 $ 
and  
$\chi^2_{SM} = 5.0 $ 
of the compatibility of the SM zero values of 
$S,\,T,\,U$ 
with the experimental data 
(\ref{eq:stue})  
at $m_H=115\,\,GeV$ 
and $m_H=300\,\,GeV$ 
respectively. 

As seen from the Fig.1 the lower limit 
  $ m^{lower}_{scalar} $  
on the masses of the scalar leptoguarks and of the scalar gluons 
is allowed by data  
(\ref{eq:stue})  
to vary within wide limits from high values 
when the contributions from these particles into 
$S,\,T,\,U$ 
are negligibly small to values of order of 
   $1\,\,TeV$ 
or less. 
It is interesting that in both cases the more light 
particles agree with the data 
(\ref{eq:stue})  
even slightly better than in the SM.  
For 
$m_H=115\,\,GeV$ 
(the curves 1) such slight improvement of the agreement 
takes place for 
$ m^{lower}_{scalar} < 400 \,\,\,GeV $  
whereas in the case of 
$m_H=300\,\,GeV$ 
 (the curves 2) such improvement is seen in all the region 
of the lightest masses of order of  
1 TeV or less and it is more appreciable also for 
$ m^{lower}_{scalar} < 400 \,\,\,GeV $. 
This improvement at 
$ m^{lower}_{scalar} < 400 \,\,\,GeV $    
takes place due to the mutual cancellation 
of the contributions into $S$ and $T$ 
from the scalar leptoquarks with those 
from the scalar gluons. 
In particular the scalar leptoquarks 
with the lightest masses of order of    
$ m^{lower}_{scalar} < 300 \,\,\,GeV $ 
(and for $\lambda_{max}=4.0 $ ) 
are compatible with the data   
(\ref{eq:stue})  
at 
$\chi^2 < 3.1 (3.2) $ 
for 
$m_H=115 (300)\,\,GeV$ 
in comparison with 
$\chi^2_{SM} = 3.5 (5.0) $ 
in the SM. 
The mass of the lightest scalar gluon in this case 
is expected to be 
$ m_{\phi_2} < 850(720) \,\,\,GeV $. 

The lightest scalar leptoquark masses 
of order  
$ m^{lower}_{scalar} \lesssim 400 \,\,\,GeV $    
are compatible with the experimental limits 
resulting from the direct search for the leptoquarks. 
The most stringent of these limits are resulted from 
the pair production and for the scalar leptoquarks 
of the first generation they give 
\cite{PDG01} 
\begin{eqnarray}
m_{LQ} > 225 \,\,GeV, \,\,\,204 \,\,GeV, \,\,\,79 \,\,GeV 
\label{eq:mlqe} 
\end{eqnarray} 
under assuming the branching ratios 
$B(eq)=1, \,\,\, 0.5, \,\,\, 0 $ 
respectively. 
It should be noted that in the model under consideration 
the coupling constants of the scalar leptoquark doublets 
with the fermions 
( and those of the scalar gluon doublets )
are proportional to the ratios of the 
fermion masses to the SM VEV 
$ \eta=246 \,\,GeV $ 
( $g_f \sim m_f / \eta$,
 the general form of this interaction can be found in 
ref.\cite{PovSm1} )
and for ordinary quarks these coupling constants are small. 
The dominant decay modes of such leptoquarks are the 
modes with heavy quarks ( predominantly with t-quark ) 
whereas the branching ratio for the first generation 
is small 
$ 0 < B(eq) \ll 0.5 $. 
So the lower experimental limit on the masses of such 
scalar leptoquarks can be near the lowest value in 
(\ref{eq:mlqe}), 
the masses of the other scalar leptoquarks are in this case 
compatible with other experimental limits 
( including those for the second and for 
the third generations ) 
resulting from the direct search for leptoquarks. 

It should be noted also that the light scalar leptoquarks 
can be also compatible with the indirect leptoquark mass limits 
resulting from the rare decays of 
$K^0_L \rightarrow \mu e$ 
type. 
Due to the smallness of the coupling constants 
of the scalar leptoquark interaction 
with d- and s- quarks 
$g_d \sim m_d / \eta \sim 2 \cdot 10^{-5}$, 
$g_s \sim m_s / \eta \sim 0.5 \cdot 10^{-3}$  
the contributions of the scalar leptoquarks into 
$K^0_L \rightarrow \mu e$ 
width 
can be sufficiently small to satisfy the stringent 
experimental limit 
$Br(K^0_L \rightarrow \mu e) < 4.7 \cdot 10^{-12}$ 
 \cite{PDG01} 
on the branching ratio of this decay,  
even for the relatively light masses of the scalar leptoquarks. 
   
Finally the data on $Z-$ physics seemingly do not exlude  
the relatively light scalar leptoquarks. In the case 
of one degenerate scalar leptoquark doublet these data give 
the lower limit on its mass of about $400 \, GeV $   
 \cite{MEG}. 

Thus, the current direct and indirect mass limits 
for leptoquarks do not exclude the relative light scalar 
leptoquark doublets considered here whereas the experimental 
data on 
$ S, \,\, T, \,\, U $ 
not only allow the existence of such particles 
but even slightly prefer them 
to have the masses of order of 
$ m^{lower}_{scalar} < 400 \,\,\,GeV $. 
Keeping in mind the sizable magnitude of 
the coupling constants of the scalar doublets with t-quark 
($g_t \sim m_t / \eta \sim 0.7,\, g^2/4\pi \sim 0.04$)  
the search for such scalar leptoquarks and scalar gluons 
in the processes with t-quarks   
at LHC is of interest.

\begin{figure}[htb]
\centerline{\psfig{file=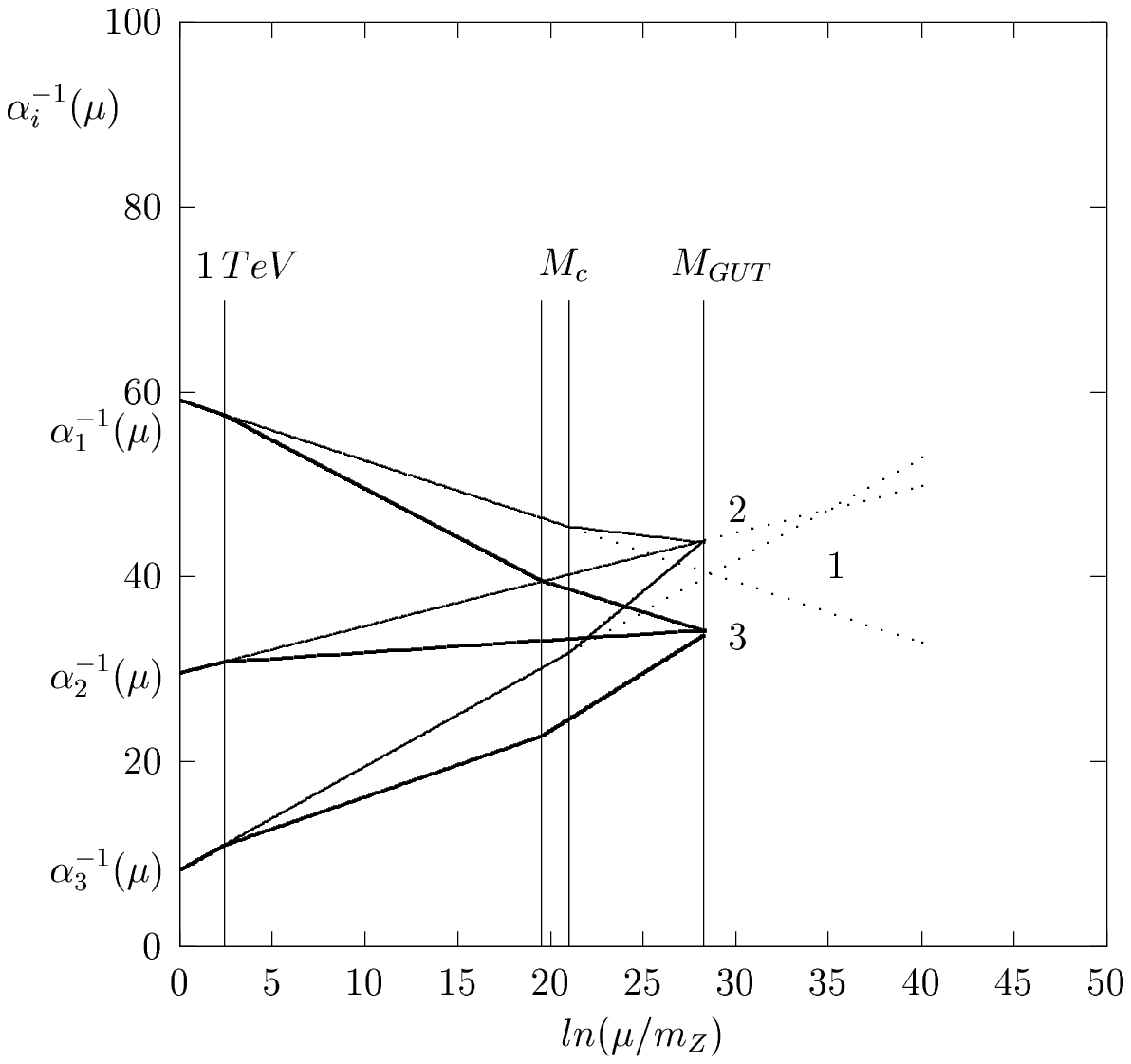,height=90mm}}
\caption{ Gauge coupling constant unification in 
        the GUT models with the four color symmetry (FCS), 
        (1)--SM, (2)--FCS without scalar doublets, 
        (3)--FCS with the light scalar doublets.  }    
\end{figure}

It should be noted that the presence of the so light 
new particles can also affect the new physics 
at high energies. In particular these particles 
can affect the gauge coupling constant unification 
in GUT approaches. 
The mass scale evolution of the running  
coupling constants $\alpha_i(\mu) = g^2_i/4\pi$, $i=1,2,3$ 
of electromagnetic (under the appropriate normalization), 
weak and strong interactions can be described in one loop 
approximation by the equations 
\begin{eqnarray}
\alpha_i^{-1}(\mu) = \alpha_i^{-1}(m_Z) - (b_i/2\pi)ln(\mu/m_Z),   
\label{eq:rcc} 
\end{eqnarray} 
where 
$b_i = b_i^{(SM)} +\Delta b_i^{(new)}$ 
are the corresponding factors of the $\beta$-functions expansion. 
In SM 
$b_i^{(SM)}=\{ 41/10,\, -19/6,\, -7 \},\, i=1,2,3$ 
and the SM without any new 
physics up to the GUT mass scale 
$M_{GUT}$  
("grand desert") do not unify three coupling constants 
at any mass scale (see the lines 1 in Fig.3). 
But such a unification can be 
possible if an intermediate new physics below   
$M_{GUT}$  
(such as the four color symmetry physics with mass scale 
$M_c$) is assumed. For example in the model under 
consideration in the case of the scalar sector 
containing, for simplicity, only the standard Higgs doublet 
and the (4,1,1) multiplet 
$\Phi^{(1)}$ 
with VEV 
$\eta_1 \sim M_c \sim 10^{11} \div 10^{12} \,\, GeV$ 
we obtain for $\mu > M_c$ 
the resulting contributions 
$\Delta b_i^{(V+\Phi^{(1)})}=\{-8/3,\, 0,\, -7/2 \}$ 
from the vector leptoquarks 
and from the multiplet $\Phi^{(1)}$ 
so that   
all three coupling constants do converge in one point at 
$M_{GUT} \sim 10^{14} \div 10^{15} \,\, GeV$ 
with 
$\alpha_3(M_{GUT}) = \alpha_2(M_{GUT}) = \alpha_1(M_{GUT}) 
\equiv  \alpha_{GUT} \sim 0.023$ 
(the lines 2 in Fig.3).  
The account of the scalar leptoquark and scalar gluon 
doublets with the masses of order of 1 TeV or less  
gives for $\mu > 1 \, TeV$
the additional contributions 
$\Delta b_i^{(S+F)} = \{37/15, \,\,  7/3, \,\, 8/3 \}$ 
so that in this case
all three coupling constants  
 $\alpha_i(\mu)$ 
do also converge in one point if
$M_c \sim 10^{10} \div 10^{11} \,\, GeV$  
and 
$M_{GUT} \sim 10^{14} \div 10^{15} \,\, GeV$ 
with
$ \alpha_{GUT} \sim 0.029  $ (the lines 3 in Fig.3). 
As seen the presence of the relatively light 
scalar leptoquark and scalar gluon doublets
(with mases below 1 TeV)
lowers the four color symmetry 
mass scale $M_c$  and increases the value of 
the unified coupling constant $\alpha_{GUT}$, 
leaving the GUT mass scale $M_{GUT}$ practically 
unchanged.  

In conclusion we can say that  
the existence of the relatively light scalar 
leptoquarks and scalar gluons 
(with masses of order of 1 TeV or less) 
is consistent with the current experimental data on S, T, U,  
the more light particles (with masses below 400 GeV) 
even slightly improve the fit. 

In particular the scalar leptoquarks   
with the masses of order of 
  $ m^{lower}_{scalar} < 300 \,\,\,GeV $  
are consistent  with current data on 
S, T, U for 
$m_H = 115(300)\,\,GeV $ 
with $\chi^2 < 3.1(3.2) $  
(in comparison with $\chi^2= 3.5(5.0) $ of the SM). 
The lightest scalar gluon in this case is expected 
to lie below $ 850(720) \,\,\,GeV. $

We emphasize the possible significance of such particles
in the top-quark physics at LHC.

{\bf Acknowledgments}


The author is grateful to 
Organizing Commitee of the International Seminar "Quarks-2002"
for possibility to participate in this Seminar.
The work was  supported 
by the Russian Foundation for Basic Research 
under grant 00-02-17883.





\begin{thebibliography}{99}
\bibitem{PS}
   J.C.~Pati and A.~Salam, Phys.~Rev. D10~(1974)~275.
\bibitem{EF}
   E.M.~Freire, Phys.~Rev. D43~(1991)~209.
\bibitem{SS}
   G.~Senjanovic, A.~Sokorac, Z.Phys. C20~(1983)~255.
\bibitem{V}
   R.R.~Volkas, Phys.~Rev. D53~(1996)~2681.
\bibitem{AD1}
   A.D.~Smirnov, Phys.~Lett. B346~(1995)~297.
\bibitem{AD2}
   A.D.~Smirnov, Yad.~Fiz. 58~(1995)~2252,
                 Phys. of At. Nucl. 58~(1995)~2137.
\bibitem{RF1}
   R.~Foot, Phys.~Lett. B420~(1998)~333.
\bibitem{RF2}
   R.~Foot, G.~Filewood, Phys. Rev.~D60~(1999)~115002.  
\bibitem{YF}
   T.L.~Yoon, R.~Foot, Phys.~Rev.~D65~(2002)~015002,hep-ph/0105101.
\bibitem{BL}
   A.~Blumhofer, B.~Lampe, Eur.~Phys.~J. C7~(1999),~141. 
\bibitem{BVR}
   W.~Buchm\"uller,R.~R\"uckl, D.~Wyler, Phys.~Lett. B191~(1987)~442. 
\bibitem{HR}
   J.L.~Hewett, T.G.~Rizzo, Phys.~Rev. D56~(1997)~5709.
\bibitem{PovSm1}
   A.V.~Povarov, A.D.Smirnov, Yad.~Fiz. 64~(2001)~78,   
                              Phys. At. Nucl. 64~(2001)~74.
\bibitem{PT}
   M.E.~Peskin and T.~Takeuchi, Phys.~Rev. D46~(1992)~381.
\bibitem{PDG01}
   Particle Data Group: D.E.~Groom et~al., Eur.~Phys.~J. C15~(2000)~1  
   and 2001 partial update for edition 2002 (URL:http://pdg.lbl.gov).  
\bibitem{AD3}
   A.D.~Smirnov, Phys.~Lett. B431~(1998)~119.
\bibitem{AD4}
   A.D.~Smirnov, Yad.~Fiz. 64~(2001)~367, 
                 Phys. At. Nucl. 64~(2001)~318.
\bibitem{AD5}
   A.D.~Smirnov, Phys.~Lett. B531~(2002)~237.
 \bibitem{MEG} 
   J.K.Mizukoshi, O.J.P.E'boli and M.C.Gonzale'z-Garci'a,
   Nucl.Phys. B443~(1995)~20.

\end{thebibliography}
\end{document}